\documentclass[
    ,final            
  ]
  {aipproc}

\usepackage{float}				
\usepackage{graphicx}			
\usepackage{amsmath}			
\usepackage{amssymb}			

\layoutstyle{8x11single}

\begin{document}

\title{Do Fermi-LAT observations really imply very large Lorentz factors in GRB outflows ?}

\classification{98.70.Rz}
\keywords      {$\gamma$-ray sources; $\gamma$-ray bursts; $\gamma$$\gamma$ annihilation}

\author{R. Hasco\"{e}t}{
  address={UPMC Univ Paris 06, UMR 7095, Institut d'Astrophysique de Paris, F-75014, Paris, France},
  altaddress={CNRS, UMR 7095, Institut d'Astrophysique de Paris, F-75014, Paris, France},
  email={hascoet@iap.fr}
}

\author{V. Vennin}{
  address={UPMC Univ Paris 06, UMR 7095, Institut d'Astrophysique de Paris, F-75014, Paris, France},
  altaddress={CNRS, UMR 7095, Institut d'Astrophysique de Paris, F-75014, Paris, France},   
  email={vennin@iap.fr}
}

\author{F. Daigne\thanks{Institut Universitaire de France}~}{
  address={UPMC Univ Paris 06, UMR 7095, Institut d'Astrophysique de Paris, F-75014, Paris, France},
  altaddress={CNRS, UMR 7095, Institut d'Astrophysique de Paris, F-75014, Paris, France},
  email={daigne@iap.fr}
}

\author{R. Mochkovitch}{
  address={UPMC Univ Paris 06, UMR 7095, Institut d'Astrophysique de Paris, F-75014, Paris, France},
  altaddress={CNRS, UMR 7095, Institut d'Astrophysique de Paris, F-75014, Paris, France},
  email={mochko@iap.fr}
}

\begin{abstract}
 Recent detections of GeV photons in a few GRBs by Fermi-LAT have led to strong constraints on the bulk Lorentz factor in GRB outflows. To avoid a large $\gamma$$\gamma$ optical depth, minimum values of the Lorentz factor are estimated to be as high as 800-1200 in some bursts. Here we present a detailed calculation of the $\gamma$$\gamma$ optical depth taking into account both the geometry and the dynamics of the jet. In the framework of the internal shock model, we compute lightcurves in different energy bands and the corresponding spectrum and we show how the limits on the Lorentz factor can be significantly lowered compared to previous estimates. 
\end{abstract}

\maketitle


\section{Introduction}


\textbf{The compactness problem.} 
The short time scales observed in GRBs (down to a few ms) can be used to deduce an upper limit on the size of the emitting region producing $\gamma$-rays. This information combined with the huge isotropic $\gamma$-ray luminosities deduced from the measured redshifts imply huge photon densities. Then the simplest assumption of an emission produced by a plasma radiating isotropically with no macroscopic motion predicts that $\gamma$-ray photons should not escape due to $\gamma$$\gamma$ annihilation $\gamma \gamma \rightarrow e^+ e^-$. This is in contradiction with the observed GRB spectra which are non-thermal and extend well above the rest-mass electron energy $m_e c^2 \approx 511$ keV.
Observation and theory can be reconciled by assuming that the emitting material is moving at ultra-relativistic velocities \cite{rees:1966}. This is mainly due to relativistic beaming.  First the relativistic beaming implies that the observer will see only a small fraction of the emitting region: the constraint on the size of the emitting region is now less severe. Second the collimation of photons in the same direction reduce the number of potential interactions. Finally the typical $\gamma$$\gamma$ interaction angle becoming small the photon energy threshold for pair production becomes higher. This theoretical context combined with the observational data gives the possibility to estimate a minimum Lorentz factor $\Gamma_{\mathrm{min}}$ for the emitting outflow in 
GRBs \cite{lithwick:2001} (or directly a Lorentz factor estimate if the $\gamma$$\gamma$ cutoff is clearly identified in the spectrum, see \cite{bregeon:2011}).
\newline \newline
\noindent \textbf{Severe constraints on the Lorentz factor from Fermi-LAT observations.} 
Since the launch of Fermi in June 2008, the LAT instrument has detected high energy photons above 10 GeV in a few GRBs. The observed $\gamma$-ray spectrum often remains consistent with a Band function covering the GBM and LAT spectral ranges without any evidence of a high energy cutoff which could be identified as a signature of $\gamma \gamma \rightarrow e^+ e^-$. This extension by Fermi of the observed spectral range upper bound from 10 MeV (e.g. BATSE) to 10 GeV implies constraints on $\Gamma_{\mathrm{min}}$ which are much more severe than the ones obtained previously. In a few cases $\Gamma_{\mathrm{min}}$ has been estimated to be of the order of 1000 (for example: GRB 080916C -- $\Gamma_{\mathrm{min}}$ = 887   \cite{abdo:2009}, GRB 090510 -- $\Gamma_{\mathrm{min}}$ = 1200 \cite{ackermann:2010}). These extreme values put severe constraints on the physics of the central engine which should be able to strongly limit the baryon load in the outflow. 

\noindent However these $\Gamma_{\mathrm{min}}$ values were obtained from a simplified ``single zone'' model where the spatial and temporal dependencies are averaged out. The motivation of this work is to develop a detailed approach taking  into account a more realistic treatment of the dynamics.

\section{Computing the $\gamma$$\gamma$ optical depth}
\vspace*{-1ex}

\textbf{General $\gamma$$\gamma$ opacity formula.} 
The $\gamma$$\gamma$ opacity ($\tau_{\gamma \gamma}$) is given by:

\begin{equation}
 \tau_{\gamma \gamma} (E_{GeV}) = \int_{l_e}^{\infty} dl \int d \Omega \int_{E_c(E_{GeV}, \psi)}^{\infty} 
 dE \ n_{\Omega}(E) \sigma_{\gamma \gamma}(E,\psi) (1-cos\psi)
\label{tau_gamma_general}
\end{equation}

\noindent All the physical quantities are measured in the laboratory (or source) frame.
$E_{GeV}$ is the energy of the photon for which $\tau_{\gamma \gamma}$ is calculated whereas $E$ is the energy of the interacting field photon.
$\psi$ represents the interaction angle between the GeV photon and the interacting photon and $\sigma_{\gamma \gamma}$ is the $\gamma$$\gamma$ interaction cross-section between these two photons. 
$E_c = 2(m_e c^2)^2/[E_{GeV}(1-cos\psi)]$ is the energy threshold of the field photon above which $\gamma$$\gamma$ annihilation can happen.
Finally $n_{\Omega}$ is the photon field distribution $[\mathrm{ph} \cdot \mathrm{cm}^{-3} \cdot \mathrm{erg}^{-1} \cdot \mathrm{sr}^{-1}]$ at a given location and time.

\noindent The equation (\ref{tau_gamma_general}) is made of a triple integral : the $dl$-integration is done over the path of the GeV photon from its emission location to the observer, the $d \Omega$-integration is done over the solid angle distribution of the interacting photon field surrounding the GeV photon whereas the $d E$-integration is done over its energy distribution. The equation (\ref{tau_gamma_general}) is general and can be applied to any photon emitted at a given location and time with a given propagation direction in the GRB outflow.  \newline

\begin{figure}[t]
\begin{tabular}{cc}
\includegraphics[scale=0.26]{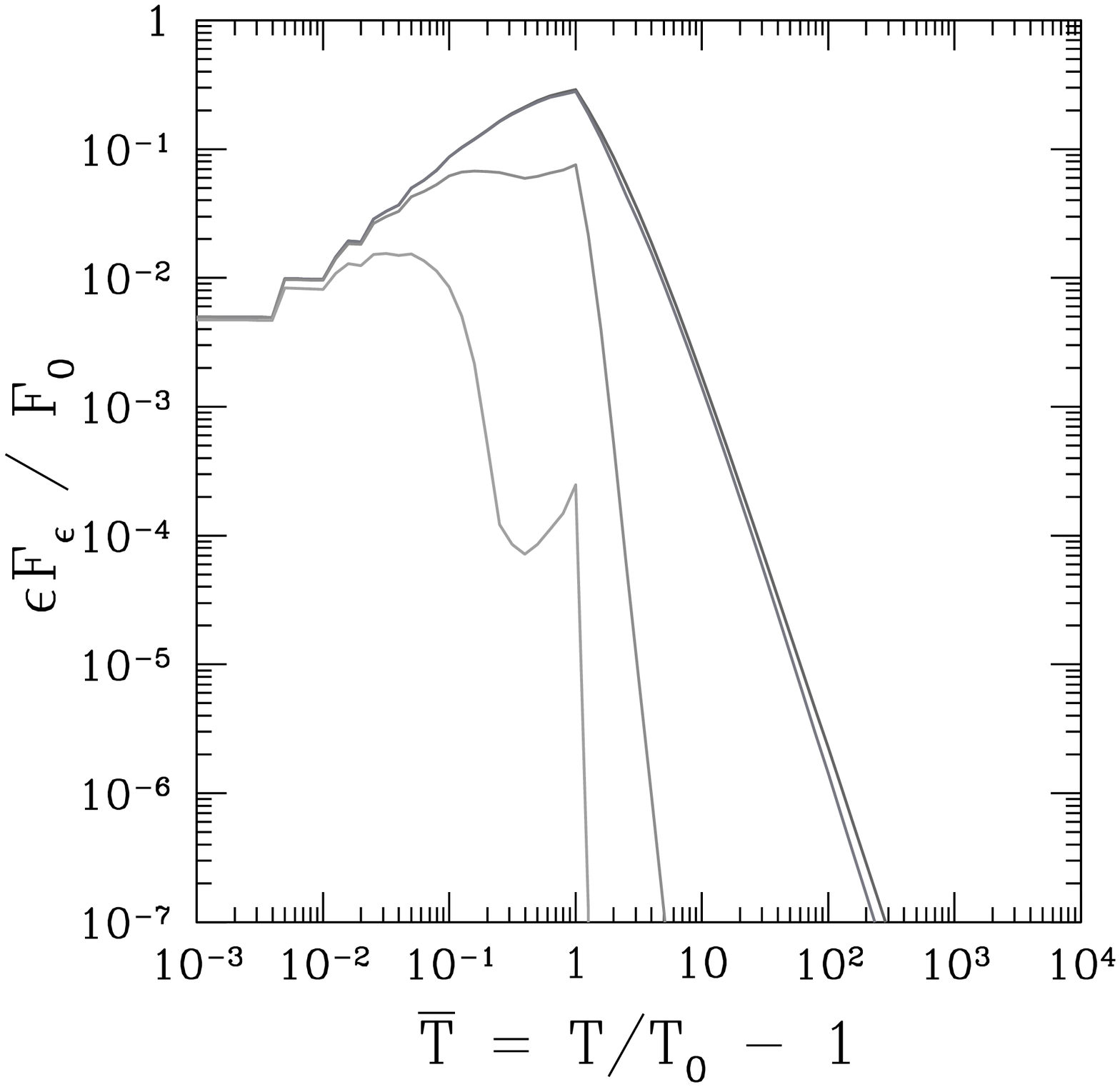} & \includegraphics[scale=0.26]{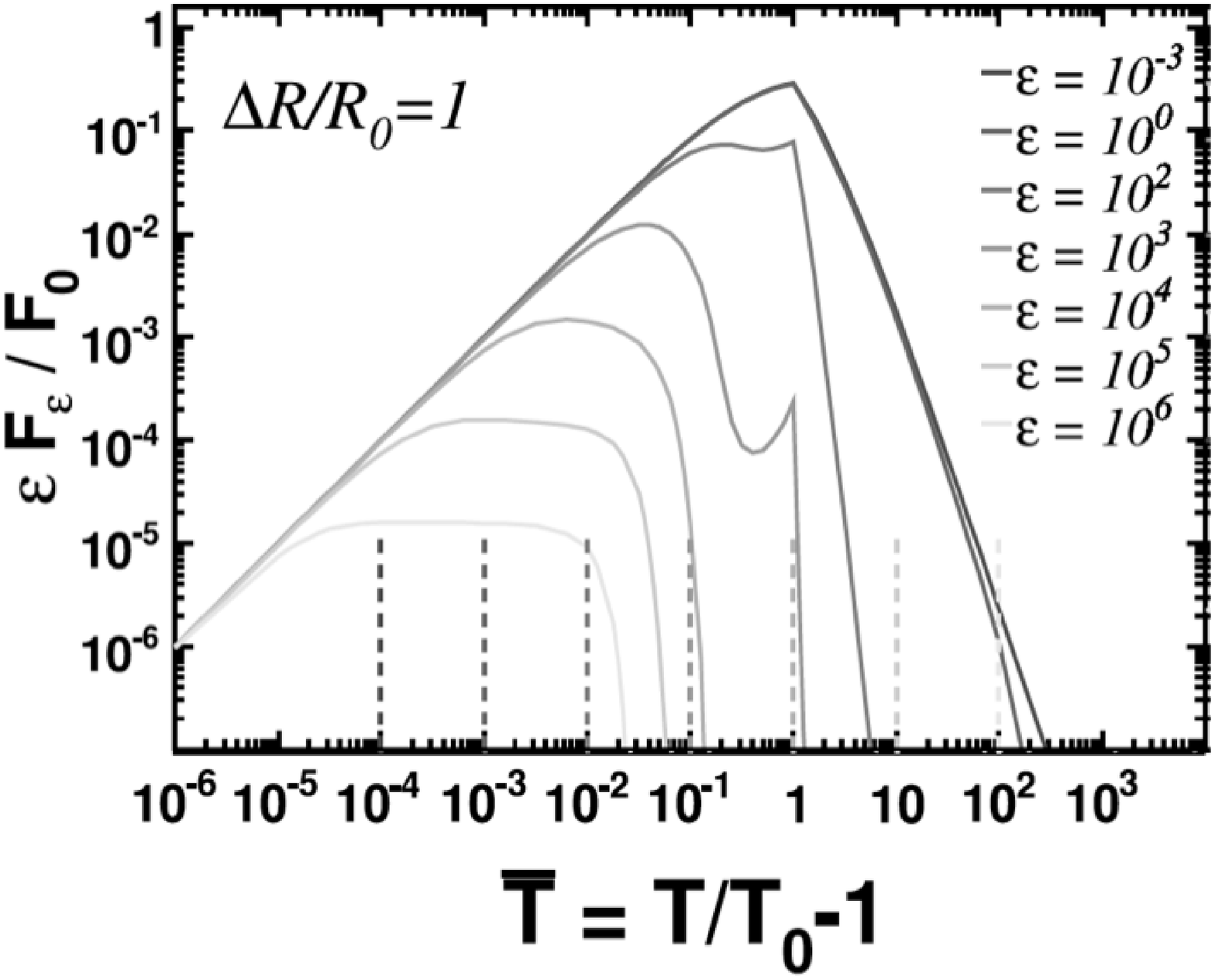} 
\end{tabular}
\caption{\textbf{Opacity in a single pulse -- comparison with the semi-analytical work by \cite{granot:2008}.}  Left panel: our model. Right panel: model presented in \cite{granot:2008}. As illustrated here, our numerical model can reproduce the normalized light-curves shown in \cite{granot:2008} (the observer time $\overline{T}$ and observed fluxes $\epsilon F_\epsilon / F_0$  have normalized values). The discrepancy for $\overline{T} < 10^{-2}$ is due to numerical resolution limitations (this corresponds to a ``true'' observer time $t_{obs} < 10^{-4}$ s).}
\label{fig_granot}
\vspace*{-10ex}
\end{figure}

\noindent \textbf{Validation of the model.} 
The kernel of our study is the calculation of the $\gamma$$\gamma$ opacity created by a spherical flash. It is then possible to model the case of a propagating radiating spherical front (representing for example a shock wave) by the succession of many spherical flashes. One of the critical step is the exact calculation of the photon density $n_{\Omega}$ taking into account all the relativistic effects. Before dealing with more complex dynamical configurations within the internal shock framework, the validity of our approach was tested on a simple single-pulse case with a comparison to the previous semi-analytic study of \cite{granot:2008} (see fig. \ref{fig_granot}).

\vspace*{-2.5ex}
\section{Application to Internal Shocks}
\vspace*{-1.5ex}

Now the model is applied to dynamical evolutions expected in the internal shock framework, where the whole prompt $\gamma$-ray emission is produced by electrons accelerated by shock waves propagating within a relativistic variable outflow.
We model the dynamics via a multiple shell model where the successive collisions between shells mimic the propagation of shock waves \cite{daigne:1998}. Each collision produces an elementary spherical flash: the simulated light curves are the result of the sum of all flashes. 
For each high energy photon, the $\gamma$$\gamma$ opacity is computed by integrating equation (1) from its emission location to the observer taking into account the exact radiation field $n_\Omega$ produced by all the collisions in the outflow.
A previous study of the $\gamma$$\gamma$ opacity in internal shock was made by \cite{aoi:2010}. However the prescription used to compute $\tau_{\gamma \gamma}$ was still approximate, using the local physical conditions of the outflow where the high energy photon is emitted and applying them to an average formula of $\tau_{\gamma \gamma}$ (as can be found in \cite{lithwick:2001, abdo:2009, ackermann:2010}).\newline

 \begin{figure}[t]
\begin{tabular}{ccc}
\includegraphics[scale=0.26]{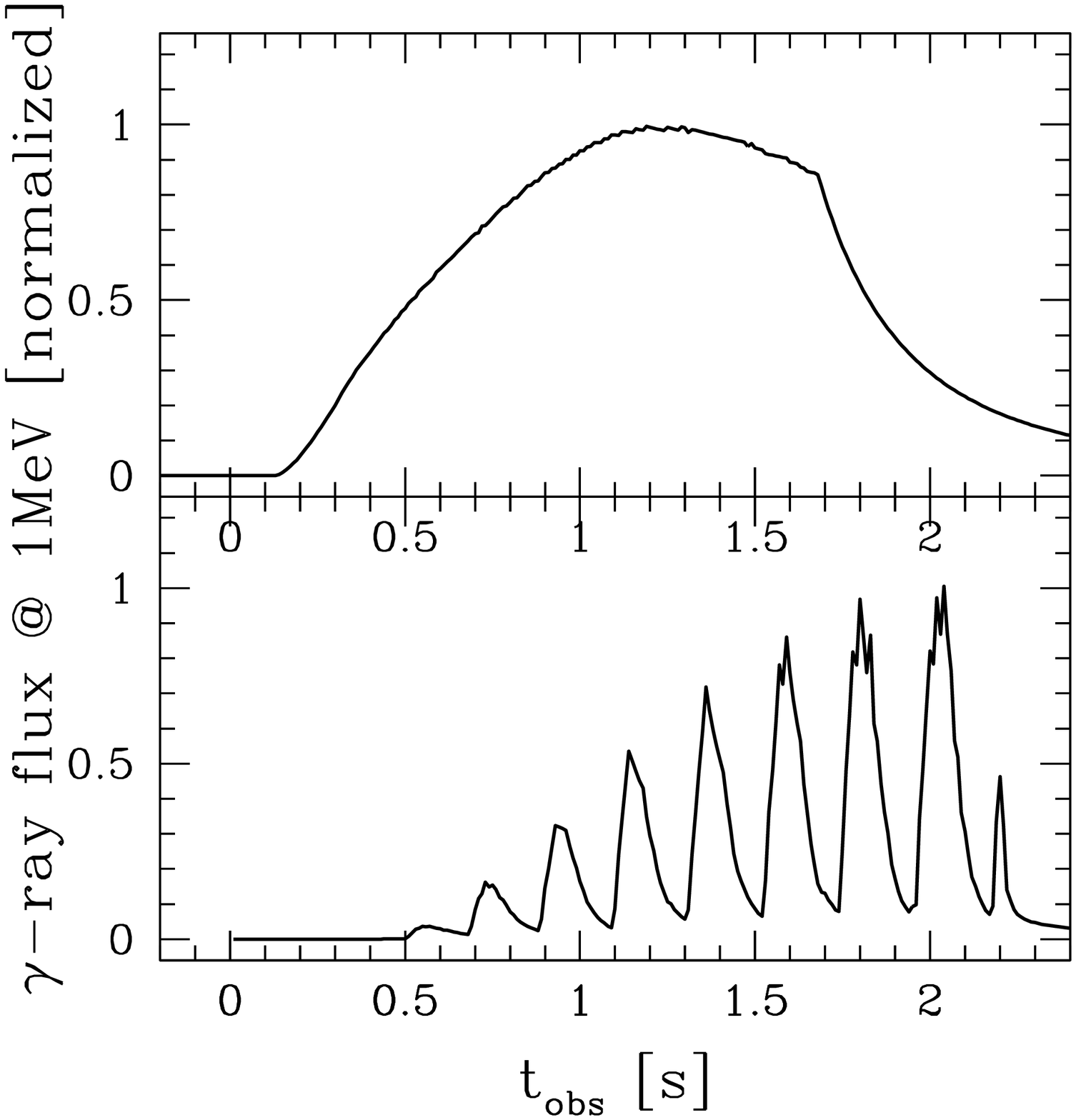} & \includegraphics[scale=0.26]{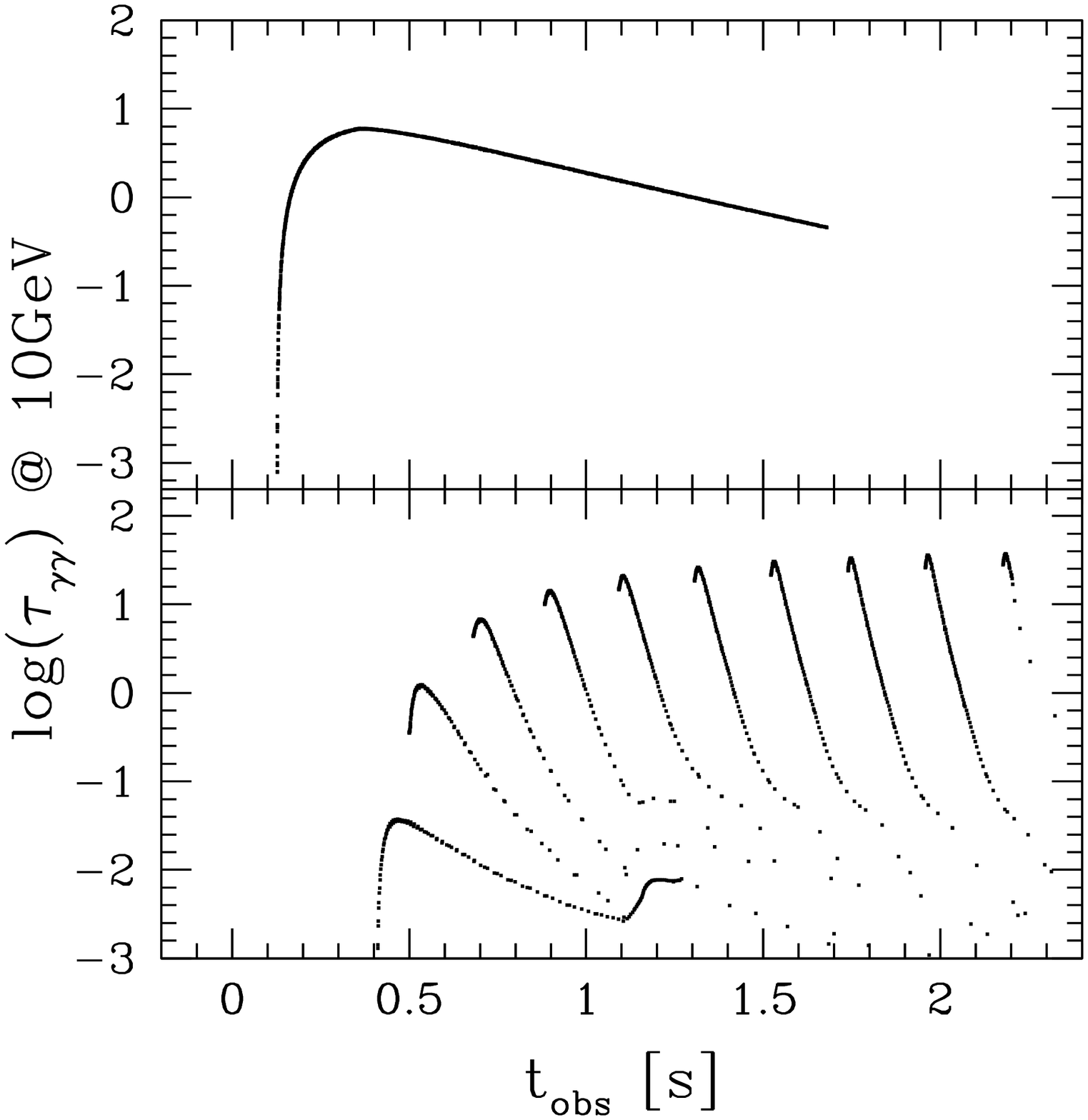} & \includegraphics[scale=0.26]{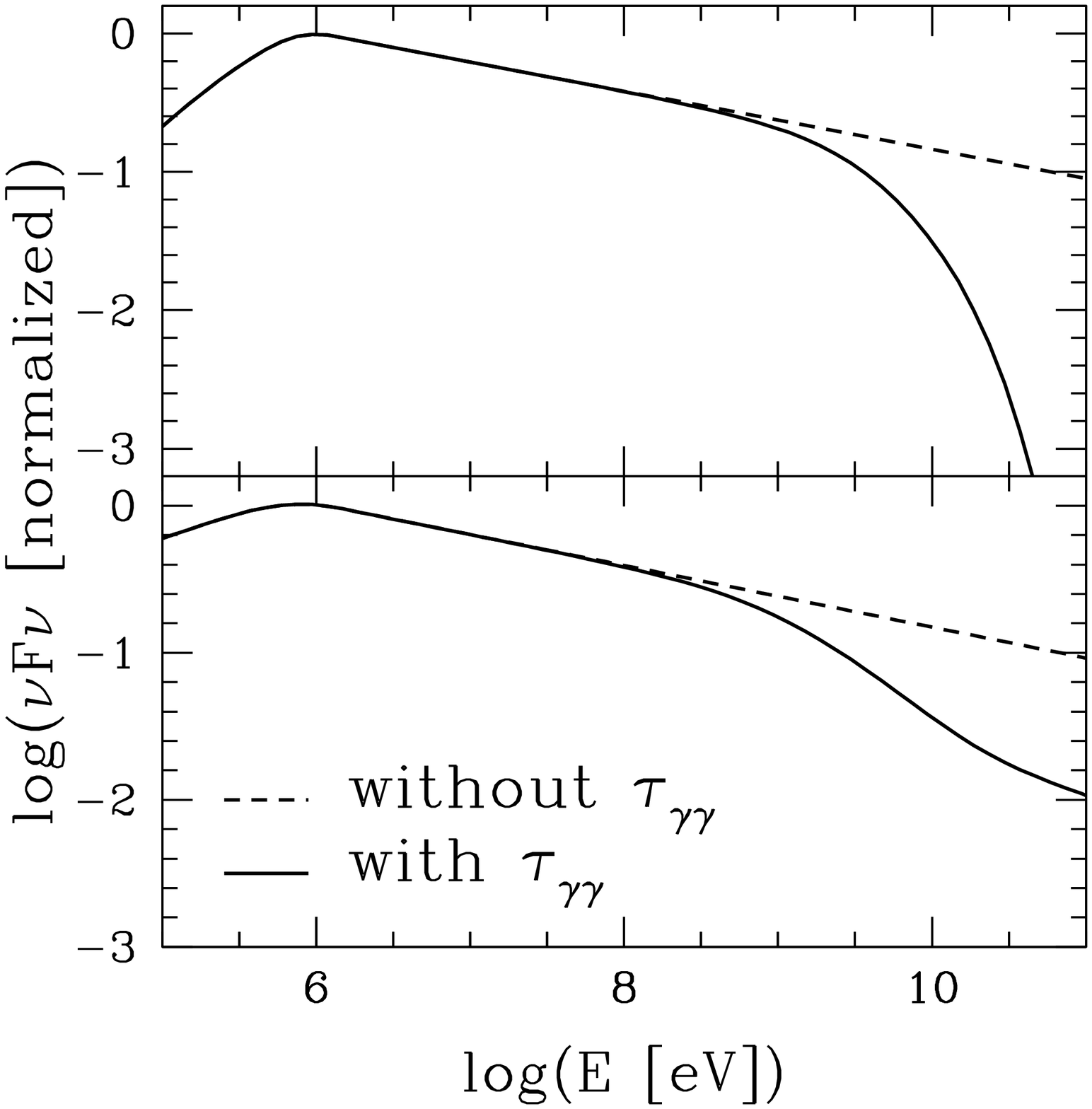}
\end{tabular}
\caption{\textbf{temporal and spectral evolution of the $\gamma$$\gamma$ opacity - Example of 2 synthetic GRBs.} The upper part of the three figures illustrates the case of a mono-pulse GRB with no additional temporal substructures whereas the bottom part show the case of a complex multiple-pulse GRB. In the first case the Lorentz factor distribution rises monotonously from 200 to 700, leading to a collision between the slow and the fast parts of the outflow. The second case is similar with the addition of a short time-scale variability. The kinetic power of the outflow is assumed to be constant. Left panel: the $\gamma$-ray light curve as a function of the observer time $\mathrm{t_{obs}}$. Middle panel:  the evolution of the $\gamma$$\gamma$ opacity seen by 10 GeV photons emitted on the line of sight as a function of the time $t_{obs}$ at which they are received by the observer. Right panel: $\gamma$-ray spectrum integrated over the GRB emission including (solid line) or not including (dashed line) the $\gamma$$\gamma$ annihilation process.}
\label{fig_scattering}
\end{figure}

\begin{figure}[h]
\begin{tabular}{cc}
\includegraphics[scale=0.26]{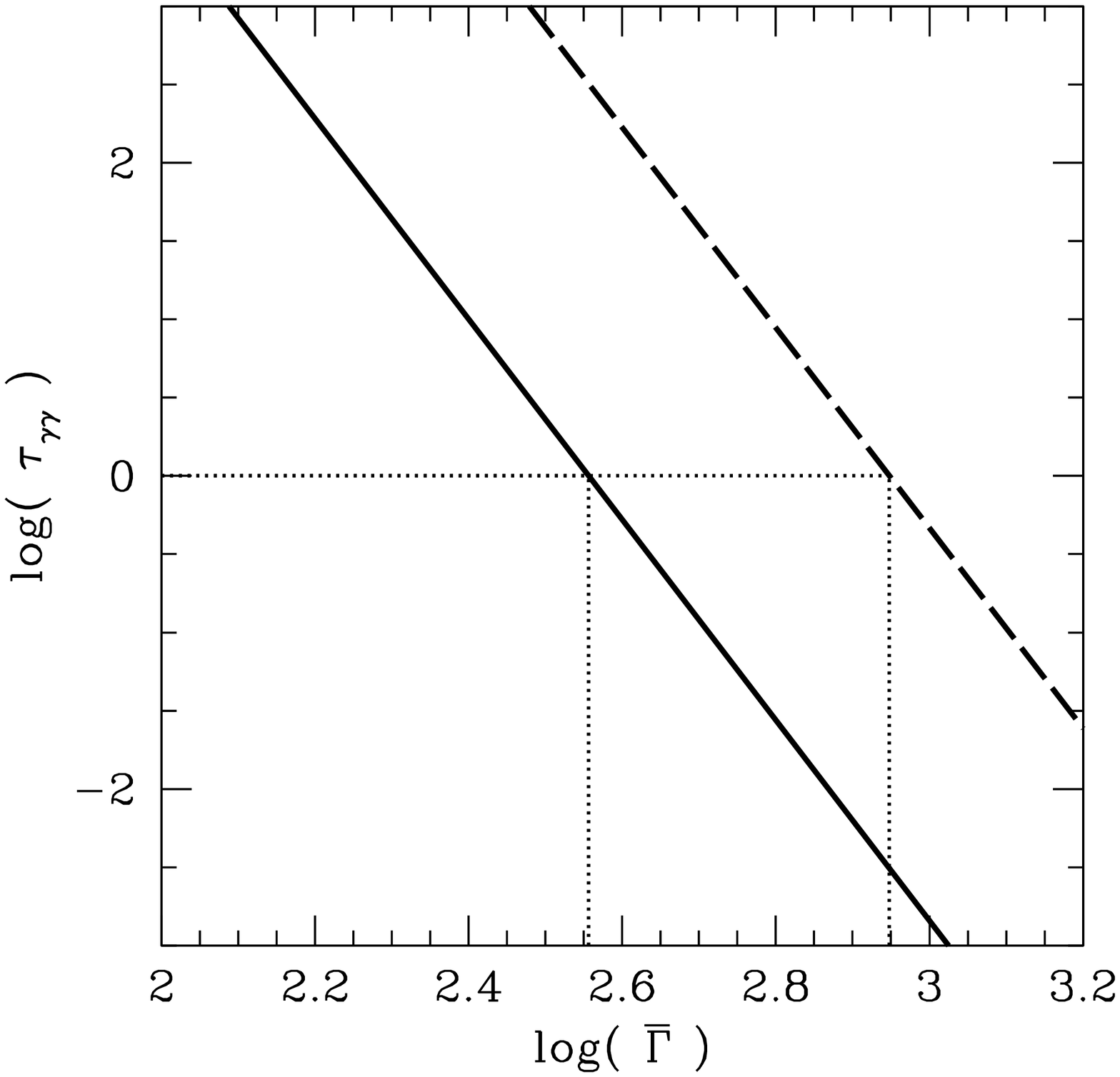} & \includegraphics[scale=0.26]{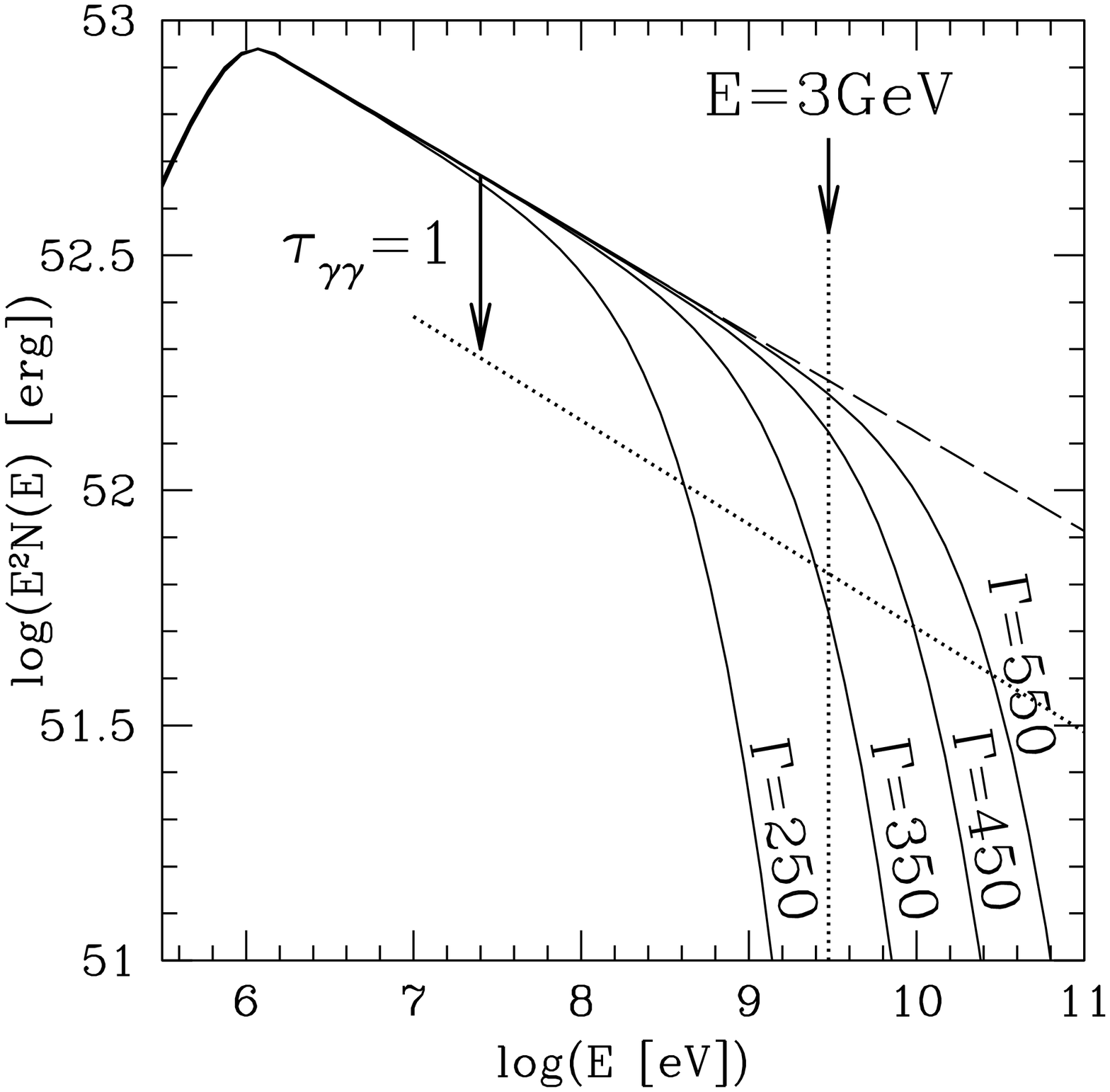}
\end{tabular}
\caption{\textbf{Minimum Lorentz factor for GRB 080916C.}  Left panel: evolution of $\tau_{\gamma \gamma}$ against the mean Lorentz factor of the outflow $\overline{\Gamma}$ following our detailed modeling (solid line) and using an average formula \cite{abdo:2009} (dashed line).  Right panel: time-integrated spectra obtained for different mean Lorentz factor of the outflow (the shape of the initial Lorentz factor distribution stays the same) and the reference spectrum without $\gamma$$\gamma$ annihilation (dashed line).}
\label{fig_gmin}
\end{figure}

\noindent \textbf{Effects of temporal and spectral evolution.} 
Due to the temporal evolution of $\tau_{\gamma \gamma}$, the opacity cutoff in a time-integrated spectrum will be smoother than a sharp exponential decay: the cutoff transition will look more like a power-law steepening. The detailed model presented here is appropriate to characterize from an observer point of view this time evolution effect.
The smoother $\tau_{\gamma \gamma}$ transition is due to the fact that the time integrated spectrum is a superposition of instant spectra which can have different $\gamma$$\gamma$ cutoff photon energies. This time evolution takes place within a same $\gamma$-ray pulse, and can be even stronger in a complex burst where the light curve is made of several pulses. Also notice that in this case cross-interactions between different pulses can become important and strongly influence the time evolution of $\tau_{\gamma \gamma}$: these potential cross-interactions are fully taken into account by the present model.
To illustrate this aspect two examples of synthetic GRBs are shown in figure \ref{fig_scattering}. In the case of the ``mono-pulse'' GRB the $\gamma$$\gamma$ opacity seen by the GRB photons remains within a decade over the whole emission (except at the very beginning when the interacting photon field progressively builds up, see also \cite{granot:2008}): the $\gamma$$\gamma$ cutoff in the time-integrated spectrum remains close to an exponential cutoff. In the case of the ``complex'' multiple-pulse GRB $\tau_{\gamma \gamma}$ covers a range of 4 decades and the  $\gamma$$\gamma$ cutoff is close to a power-law steepening. \newline

\noindent \textbf{Minimum Lorentz factor: detailed modeling vs. simple estimates.} 
One natural application of our model is the estimate of the minimum bulk Lorentz factor $\Gamma_{min}$ in GRB outflows. Our model reproduces the dependence of $\tau_{\gamma \gamma}$ on the mean Lorentz factor $\overline{\Gamma}$ of the outflow and the energy of the high energy photon $E_{GeV}$: $\tau_{\gamma \gamma} \propto \Gamma^{2(\beta-1)} E_{GeV}^{-(1+\beta)}$. However the normalization we obtain is different compared to a simple ``single zone'' model (see fig. \ref{fig_gmin}, left panel) and gives values of $\Gamma_{min}$ which are lower by a factor of 2-3 (the exact factor depending on the details of the dynamical features of the GRB). 
To illustrate this aspect with an example our approach was applied to the case of GRB 080916C. Using our numerical model a synthetic GRB was generated which reproduces the main observational characteristics of the GRB: the total radiated isotropic $\gamma$-ray energy ($E_{iso} = 8.8 \times 10^{54}$ ergs), the spectral properties ($E_p$, $\alpha$, $\beta$ parameters of the Band function), the envelop of the light curve and a short time-scale variability of 2s in the observer frame. The study is focused on the most constraining time bin (time bin b), during which the highest observed photon energy was 3 GeV (16 GeV in the source rest frame). $\Gamma_{min}$ is obtained by requiring that for this photon $\tau_{\gamma \gamma} < 1$ (see figure \ref{fig_gmin}, right panel). Following our approach we find a minimum Lorentz factor $\Gamma_{min} = 360$, instead of 887 which was obtained from an approximate ``single zone'' model \cite{abdo:2009}.

\section{Conclusions}

The detailed $\gamma$$\gamma$ opacity calculation model presented in these proceedings is appropriate and accurate to study many aspects and consequences of $\gamma$$\gamma$ annihilation in GRBs. In the present work we focus on the internal shock model and consider the consequences and signatures that $\gamma$$\gamma$ opacity could have in GRB observations. The model was validated by comparing our results to a previous semi-analytical study \cite{granot:2008}.
\begin{itemize}
\item It is shown how a detailed calculation can predict minimum Lorentz factors $\Gamma_{min}$ which are lower by a factor of 2-3 compared to a simplified ``single zone'' model where spatial and temporal dependencies are averaged out. 
\item The $\gamma$$\gamma$ cutoff transition is also characterized in time-integrated spectra. It is usually closer to a power-law steepening than to a sharp exponential cutoff. The exact shape of the transition strongly depends on the details of the GRB dynamics.
\end{itemize}
Other effects (Hascoët et al. in prep.) can be studied with our model: 
\begin{itemize}
\item The temporal evolution of $\tau_{\gamma \gamma}$ during a burst could favor a delay between the MeV and GeV light curves.
\item For complex GRBs, the $\gamma$$\gamma$ opacity could suppress the shortest time-scale features in high energy light curves (above 100 MeV).
\item If MeV and GeV photons are not produced at the same location (see e.g. \cite{bosnjak:2009,zou:2011}), $\tau_{\gamma \gamma}$ could be further lowered, reducing even more the constraint on $\Gamma_{min}$.
 \end{itemize}



\begin{theacknowledgments}
This work is partially supported by a grant from the French Space Agency (CNES).\newline
R.H. is funded by the research foundation from ''Capital Fund Management''.
\end{theacknowledgments}

\bibliographystyle{aipproc}   
\bibliography{hascoet}

\end{document}